 \documentclass[final,5p,times,twocolumn,sort&compress]{elsarticle}

\usepackage{graphicx}
\usepackage{amssymb}
\usepackage{amsthm}
\usepackage{bm} 
\usepackage{amsthm}
\usepackage{latexsym}
\usepackage{float}
\usepackage{amsfonts}
\usepackage{amsmath}
\usepackage{amssymb}
\usepackage{color,graphicx}

\journal{a journal}

\begin{document}

\begin{frontmatter}



\title{Correlation and dephasing effects on the non-radiative coherence between bright excitons in an InAs QD ensemble measured with 2D spectroscopy}


\author[JILA,CU]{G. Moody}
\author[JILA,CU]{R. Singh}
\author[JILA]{H. Li}
\author[Dortmund,Ioffe]{I. A. Akimov}
\author[Dortmund]{M. Bayer}
\author[Bochum]{D. Reuter}
\author[Bochum]{A. D. Wieck}
\author[JILA,CU]{S. T. Cundiff\corref{cor1}}
\address[JILA]{JILA, University of Colorado \& National Institute of Standards and Technology, Boulder CO 80309-0440}
\address[CU]{Department of Physics, University of Colorado, Boulder CO 80309-0390}
\address[Dortmund]{Experimentelle Physik 2, Technische Universit$\ddot{\textit{a}}$t Dortmund, D-44221 Dortmund, Germany}
\address[Ioffe]{A. F. Ioffe Physical-Technical Institute, Russian Academy of Sciences, 194021 St. Petersburg, Russia}
\address[Bochum]{Lehrstuhl fuer Angewandte Festkoerperphysik, Ruhr-Universitaet Bochum, Universitaetsstrasse 150, D-44780 Bochum, Germany}
\cortext[cor1]{Corresponding Author. \\    \textit{Email Address}: cundiff@jila.colorado.edu (S. T. Cundiff)}

\begin{abstract}
Exchange-mediated fine-structure splitting of bright excitons in an ensemble of InAs quantum dots is studied using optical two-dimensional Fourier-transform spectroscopy.  By monitoring the non-radiative coherence between the bright states, we find that the fine-structure splitting decreases with increasing exciton emission energy at a rate of 0.1 $\mu$eV$/$meV.  Dephasing rates are compared to population decay rates to reveal that pure dephasing causes the exciton optical coherences to decay faster than the radiative limit at low temperature, independent of excitation density.  Fluctuations of the bright state transition energies are nearly perfectly-correlated, protecting the non-radiative coherence from interband dephasing mechanisms.
\end{abstract}


\end{frontmatter}


\section{Introduction}
\label{Introduction}

In semiconductor quantum dots (QDs), the lowest energy electron-hole pair (exciton) has a coherence time up to nanoseconds \cite{Langbein2004}, making QDs promising candidates for forming the building blocks of devices relevant for quantum information \cite{Bonadeo1998,Chen2000,Stievater2001,Li2003,Mar2010} and nonlinear electro-optic applications \cite{Englund2012}.  Without considering the exchange interaction, the lowest-energy exciton state is four-fold spin degenerate, with two optically active ($\left|\pm1\right>$) and inactive ($\left|\pm2\right>$) states.  Relative to higher-dimensional quantum systems, confinement in QDs increases the strength of the exchange interaction, which couples the spins of the electron and hole \cite{Gammon1996,Bayer2002a}.  Through short-range exchange, the four-fold degeneracy is lifted and the bright $\left|\pm1\right>$ states are energetically split from the dark $\left|\pm2\right>$ states.  In QDs that lack cylindrical symmetry of the confinement potential, the long-range exchange interaction couples the $\left|+1\right>$ and $\left|-1\right>$ states to form two orthogonal linearly-polarized exciton states, $\left|\textrm{H}\right>$ and $\left|\textrm{V}\right>$, that are energetically separated by the so-called fine-structure splitting, $\delta_{\textrm{1}}$.  These states are coupled by confinement-enhanced Coulomb interactions, forming a four-level diamond system shown in Fig. \ref{fig1}.

Successful implementation of the aforementioned applications using QDs requires understanding how the physical properties of the QD -- including confinement, internal strain and alloying -- affect its optical properties.  Significant insight in this regard has been obtained in the past decade, either through single QD studies using photoluminescence \cite{Bayer2002a,Tartakovskii2004,Seguin,Greilich2006,Abbarchi2008} or by spectrally-integrating the ensemble nonlinear response using time-integrated four-wave mixing (FWM) techniques \cite{Langbein2004a}.  These studies show that $\delta_{\textrm{1}}$ typically decreases with increasing emission energy (decreasing confinement) at a rate ranging from 0.25 to 2 $\mu$eV$/$meV.  Anisotropy in the QD confinement potential responsible for $\delta_{\textrm{1}}$ tends to orient $\left|\textrm{H}\right>$ and $\left|\textrm{V}\right>$ along the [110] and [1$\bar{\textrm{1}}$0] crystal axes for samples grown along [001].  Elongation of the QD shape occurring during growth and strain-induced piezoelectric fields \cite{Seguin} are thought to be responsible for the anisotropy, although the dominant mechanism is still under debate \cite{Abbarchi2008}.

In this work, we demonstrate that $\delta_{\textrm{1}}$ can be measured for all QDs in the inhomogeneously broadened ensemble \textit{simultaneously} by monitoring the temporal evolution of the coherence \cite{Yang2008a} between $\left|\textrm{H}\right>$ and $\left|\textrm{V}\right>$ using two-dimensional Fourier-transform spectroscopy (2DFTS), which is based on three-pulse transient FWM \cite{Cundiff2012}.  The non-radiative $\left|\textrm{H}\right>$--$\left|\textrm{V}\right>$ coherence, which has no optical dipole moment, can be probed optically by coherently exciting the $\left|\textrm{0}\right>$--$\left|\textrm{H}\right>$ and $\left|\textrm{0}\right>$--$\left|\textrm{V}\right>$ transitions using either two orthogonal linearly-polarized pulses or a single circularly-polarized pulse.  Previous experiments \cite{Chen2000,Gammon2001,Li2004,Poem2011} have used this coherence to manipulate exciton spin states and create entangled photon pairs from single QDs.  Here we measure the dynamics of this coherence in an InAs QD ensemble and present zero-quantum 2D spectra that reveal the effects of dephasing and correlated scattering on the exchange-split exciton states.  We find that within the full-width half-maximum (FWHM) of the inhomogeneous distribution, $\delta_{\textrm{1}}$ decreases with increasing emission energy at a rate of 0.1 $\mu$eV$/$meV and anisotropy in the QD confinement potential responsible for $\delta_{\textrm{1}}$ is oriented along the same crystal axis for all QDs.  By comparing the exciton homogeneous linewidth \cite{Borri2001}, comprised of a narrow Lorentzian zero-phonon line (ZPL) at low temperatures with a broad phonon background at high temperatures, to the population decay rate, we find that additional pure dephasing broadens the linewidth beyond the radiative limit at low temperature, independent of excitation density.  Interestingly, the dephasing rate of the $\left|\textrm{H}\right>$--$\left|\textrm{V}\right>$ coherence is equal to the population decay rate of the individual exciton states, indicating that nearly perfectly-correlated scattering of $\left|\textrm{H}\right>$ and $\left|\textrm{V}\right>$ shield the non-radiative coherence from the pure dephasing mechanisms.

\section{Sample and Technique}
\label{Sample and Technique}
\subsection{Self-Assembled InAs Quantum Dots}
\label{Sample}

\begin{figure}[h]
\centering
\includegraphics[width=0.95\columnwidth]{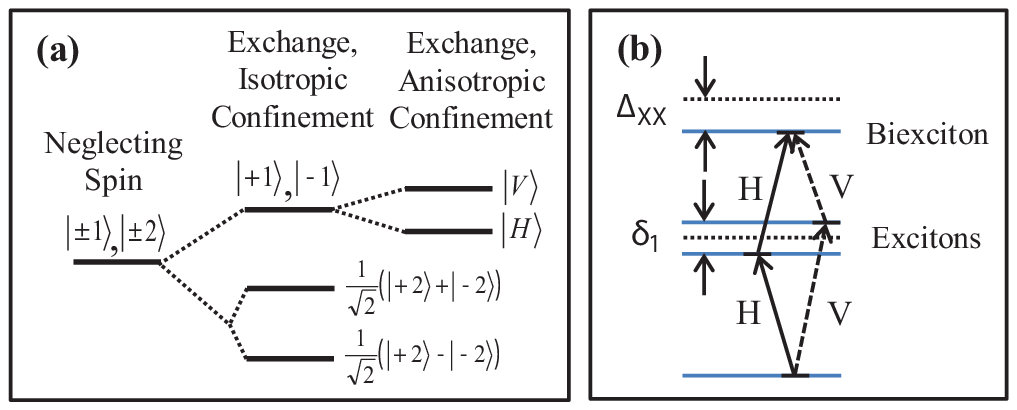}
\caption{(Color online) (a) Neglecting spin, the $\left|\pm1\right>$ and $\left|\pm2\right>$ exciton spin states are degenerate.  When considering electron-hole exchange in an isotropic confinement potential, the $\left|\pm1\right>$ states are energetically split from the $\left|\pm2\right>$ states, which are coupled.  In an anisotropic confinement potential, the $\left|+1\right>$ and $\left|-1\right>$ states couple to form a basis of two orthogonal linearly-polarized exciton states, $\left|\textrm{H}\right>$ and $\left|\textrm{V}\right>$, that are separated by the so-called fine-structure splitting, $\delta_{\textrm{1}}$.  (b) An energy level scheme is shown for a neutral QD, indicating $\left|\textrm{H}\right>$ and $\left|\textrm{V}\right>$ separated by $\delta_{\textrm{1}}$ and the biexciton state with biexciton binding energy $\Delta_{\textrm{XX}}$.}
\label{fig1}
\end{figure}

The sample consists of ten quantum-mechanically isolated layers of self-assembled InAs QDs with GaAs barriers epitaxially grown on a GaAs (001) substrate.  The sample is thermally annealed after growth at 900 $^{\circ}$C for 30 seconds, which blue-shifts the peak absorption to $\sim$ 1345 meV, narrows the inhomogeneous linewidth to $\sim$ 15 meV FWHM and decreases the ground state-to-wetting layer confinement to $\sim$ 100 meV.  Impurities unintentionally introduced during growth result in approximately half of the QDs being charged with a hole, determined through independent single QD studies and a quantitative analysis discussed elsewhere \cite{Moody2012}.  Charged QDs exhibit no fine-structure splitting because an optically-excited electron-hole pair binds with the resident hole to form a trion in which the holes are in a singlet state that is insensitive to exchange \cite{Bayer2002a}.  The sample is held in a liquid helium cold-finger cryostat at 10 K.  An energy level scheme is shown in Fig. \ref{fig1}(b) for a neutral QD.  By taking advantage of the polarization selection rules and scanning specific pulse delays, we can unambiguously measure the exciton and trion dephasing rates, population decay rates and the $\left|\textrm{H}\right>$--$\left|\textrm{V}\right>$ coherence dephasing rate.

\subsection{Optical 2D Fourier-Transform Spectroscopy}
\label{Technique}

\begin{figure}[h]
\centering
\includegraphics[width=1.0\columnwidth]{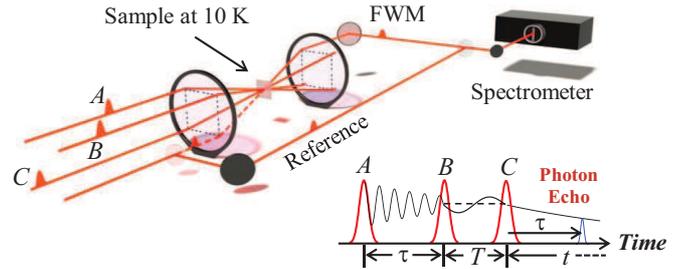}
\caption{(Color online) Three pulses are incident on the sample to generate a FWM signal, which is heterodyned with the phase-stabilized reference pulse and spectrally-resolved.  The sample is held in a liquid helium cold-finger cryostat at 10 K.  The delay $\tau$ ($T$) is stepped while $T$ ($\tau$) is held fixed at 200 fs, and interferograms are recorded at each delay.  The data are numerically Fourier-transformed with respect to $\tau$ ($T$) to generate a rephasing one-quantum (zero-quantum) spectrum.}
\label{fig2}
\end{figure}

2DFTS has proven to be a powerful tool for investigating QD ensembles.  The technique has facilitated our understanding of exciton dephasing \cite{Moody2011b} and relaxation dynamics \cite{Moody2011}, exciton-exciton coherent coupling \cite{KasprzakJ.2011} and electronic properties \cite{Harel2012}.  In this work, 2DFTS is used to study exciton fine-structure splitting with a spectral resolution determined by the maximum delay between pulses, allowing us to achieve up to 30 neV resolution.  The experimental setup, shown in Fig. \ref{fig2}, is based on three-pulse transient FWM with the addition of interferometric stabilization of the pulse delays \cite{Bristow2009}.  150-fs pulses with wave vectors $\textbf{\textit{k}}_{\textit{a}}$, $\textbf{\textit{k}}_{\textit{b}}$ and $\textbf{\textit{k}}_{\textit{c}}$ are focused onto the sample to generate a FWM signal in the direction $\textbf{\textit{k}}_{\textit{s}}=-\textbf{\textit{k}}_{\textit{a}}+\textbf{\textit{k}}_{\textit{b}}+\textbf{\textit{k}}_{\textit{c}}$.  The excitation spectrum is tuned to be resonant with the inhomogeneously-broadened ground state distribution and has a FWHM of $\sim$ 10 meV.  The signal is interfered with a phase-stabilized reference pulse and their interference is spectrally resolved with a resolution of $\sim$ 17 $\mu$eV.  Pulse $A$ is incident on the sample first, followed by pulse $B$ after a time $\tau$.  Pulse $C$ arrives after pulse $B$ with a delay $T$, and the signal is emitted during $t$.  Interferograms are recorded while scanning $\tau$ ($T$) and holding $T$ ($\tau$) fixed at 200 fs.  At each delay, the phases of pulses $A$ and $B$ are toggled between zero and $\pi$ by liquid crystal modulators and phased interferograms are added appropriately to eliminate scatter of the excitation pulses along $\textbf{\textit{k}}_{\textit{s}}$.  The excitation intensity is fixed at 65 W cm$^{-2}$ (3 $\times$ $10^{12}$ photons $\cdot$ pulse$^{-1}$ $\cdot$ cm$^{-2}$), for which the signal is maximum while remaining in the $\chi^{\left(\textrm{3}\right)}$ regime.  At this excitation density, we estimate that an average of 0.03 excitons are excited per QD.  The beam polarizations can be independently aligned along the [110] and [1$\bar{\textrm{1}}$0] crystal axes, which we define as H and V, and a half-wave plate and polarizer are placed before the spectrometer to select either the H or V component of the FWM signal.

\section{Results and Discussion}
\label{Results and Discussion}

\begin{figure*}[t]
\centering
\includegraphics[width=2\columnwidth]{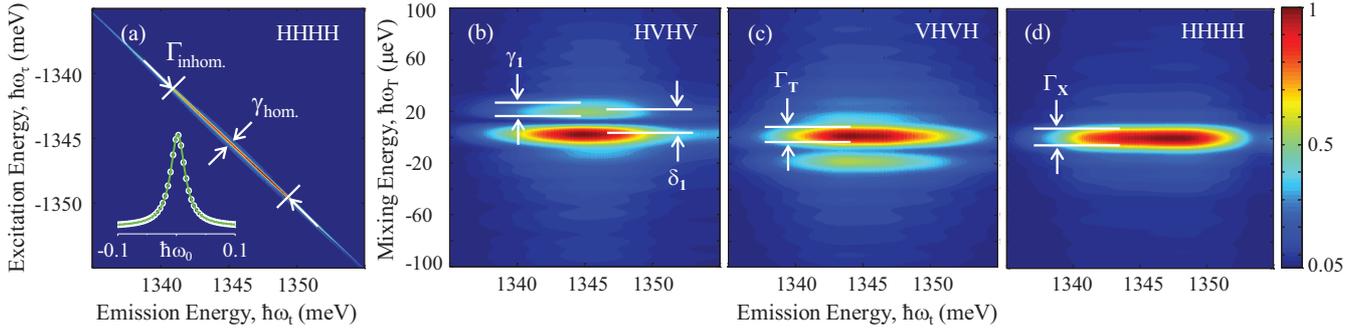}
\caption{(Color online) (a) Rephasing one-quantum amplitude spectrum for HHHH polarization.  A single peak is observed inhomogeneously-broadened along the diagonal ($\gamma_{\textrm{inhom.}}$) and homogeneously-broadened along the cross-diagonal ($\gamma_{\textrm{hom.}}$). The inset shows a cross-diagonal slice (points) taken at $\hbar\omega_t = 1345$ meV and (Lorentzian)$^{1/2}$ fit (line) plotted as a function of emission energy. (b)-(d) Rephasing zero-quantum amplitude spectra for HVHV, VHVH and HHHH polarization sequences, respectively.  For HVHV and VHVH polarizations, the spectra feature an $\left|\textrm{H}\right>$--$\left|\textrm{V}\right>$ coherence peak at $+\delta_{\textrm{1}}$ and $-\delta_{\textrm{1}}$ along $\hbar\omega_T$, respectively, and a trion peak at $\hbar\omega_T = 0$ $\mu$eV.  Only a single peak dominated by the excitonic nonlinear response is observed for HHHH polarization.}
\label{fig3}
\end{figure*}

A rephasing one-quantum amplitude spectrum, shown in Fig. \ref{fig3}(a), is generated by scanning $\tau$ and taking a numerical Fourier transform of the extracted FWM signal with respect to this delay.  A co-linear (HHHH) polarization scheme is used, where the polarization sequence is defined as that of pulses $A$, $B$, $C$ and the detected signal, respectively.  Because pulse $A$ is conjugated and is incident on the sample first, the signal oscillates at negative frequencies during $\tau$ with respect to oscillations during $t$; therefore we plot the spectrum as a function of negative excitation energy, $-\hbar\omega_{\tau}$, and positive emission energy, $\hbar\omega_{t}$, along the vertical and horizontal axes, respectively.  Excitons and trions are optically-accessible using HHHH polarization and spectrally overlap along the diagonal due to inhomogeneity \cite{Siemens2010} from a distribution of QD sizes.  For these states,  pulse $A$ creates a first-order optical coherence along the $\left|\textrm{0}\right>$--$\left|\textrm{H}\right>$ transition.  After a time $\tau$, pulse $B$ converts this coherence to either a ground- or excited-state population grating.  After a time $T$, pulse $C$ generates a coherence along the $\left|\textrm{0}\right>$--$\left|\textrm{H}\right>$ transition that radiates along $\textbf{\textit{k}}_{\textit{s}}$.  An off-diagonal biexciton peak is expected but is too weak to be observed in the spectrum with these excitation conditions.

The ZPL width, $\gamma$, related to the population decay rate, $\Gamma$, and the pure dephasing rate, $\gamma^*$, by $\gamma = \Gamma/2 + \gamma^*$, is determined by fitting \cite{Siemens2010} a cross-diagonal slice to a single (Lorentzian)$^{1/2}$, shown in the inset to Fig. \ref{fig3}(a), since at low temperature the ZPL is observed without a broad phonon background.  In previous work, we found that the exciton dominates for HHHH, whereas when using an HVVH polarization sequence, no quantum-mechanical pathways exist for the exciton, and the biexciton and trion nonlinear responses are isolated \cite{Moody2012}.  From these experiments, we determined that the exciton and trion  ZPL widths, $\gamma_{\textrm{X}}$ and $\gamma_{\textrm{T}}$, are equal to 12$\pm1$ and 8$\pm2$ $\mu$eV, respectively, after deconvolving the spectrometer response, and are constant within the FWHM of the inhomogeneous distribution.  Within the estimated uncertainties, $\gamma_{\textrm{X,H}} = \gamma_{\textrm{X,V}}$.

The non-radiative coherence between states $\left|\textrm{H}\right>$ and $\left|\textrm{V}\right>$ is created by cross-polarizing pulses $A$ and $B$ and is probed by scanning $T$ instead of $\tau$.  For an HVHV polarization sequence, pulse $A$ generates an optical coherence along the $\left|\textrm{0}\right>$--$\left|\textrm{H}\right>$ transition and pulse $B$, incident on the sample at $\tau = 200$ fs to avoid coherent artifacts and pulse time-ordering ambiguities, generates an optical coherence along the $\left|\textrm{0}\right>$--$\left|\textrm{V}\right>$ transition.  The two optical coherences are equivalent to a non-radiative coherence between $\left|\textrm{H}\right>$ and $\left|\textrm{V}\right>$ that oscillates during $T$ with a frequency equal to $\delta_{\textrm{1}}/\hbar$.  Pulse $C$ converts this coherence back to an optical coherence along the $\left|\textrm{0}\right>$--$\left|\textrm{V}\right>$ transition, which radiates along $\textbf{\textit{k}}_{\textit{s}}$ and is recorded while the delay $T$ is stepped.  The signal is Fourier transformed with respect to $T$ to generate a zero-quantum spectrum, shown in Figs. \ref{fig3}(b), \ref{fig3}(c) and \ref{fig3}(d) for HVHV, VHVH and HHHH polarization sequences, respectively.

The spectrum shown in Fig. \ref{fig3}(b) features a peak at zero mixing energy that is inhomogeneously broadened along $\hbar\omega_{t}$, which we attribute to the trion nonlinear response since no quantum-mechanical pathways exist for the exciton for this polarization sequence.  Being at zero mixing energy, the trion is in either a ground- or excited-state population during $T$, and the linewidth along $\hbar\omega_{T}$ is a measure of the population decay rate.  The spectrum also features a peak at $\hbar\omega_T \approx 19$ $\mu$eV, which we attribute to the $\left|\textrm{H}\right>$--$\left|\textrm{V}\right>$ coherence that oscillates with frequency $\omega_{\textrm{VH}} = \omega_{\textrm{V}} - \omega_{\textrm{H}}$ during $T$.

The spectrum shown in Fig. \ref{fig3}(c) demonstrates that upon switching the polarization sequence to VHVH, for which during $T$ the coherence oscillates at frequency $\omega_{\textrm{HV}} = - \omega_{\textrm{VH}}$, the $\left|\textrm{H}\right>$--$\left|\textrm{V}\right>$ coherence frequency switches sign as expected.  Because this feature is a well-defined peak for a specific emission energy, we conclude that the linearly-polarized exciton states are aligned along the same crystal axes for the ensemble and the lower-energy exciton state is correctly labeled as $\left|\textrm{H}\right>$.  If the confinement anisotropy is randomly oriented, then a distribution of $\delta_{\textrm{1}}$ would be present and well-defined peaks at $\hbar\omega_T = +\delta_{\textrm{1}}$ and $\hbar\omega_T = -\delta_{\textrm{1}}$ for HVHV and VHVH polarizations, respectively, would not be observed.  When using an HHHH polarization sequence, shown in Fig. \ref{fig3}(d), the $\left|\textrm{H}\right>$--$\left|\textrm{V}\right>$ coherence peak disappears, since during $T$ only exciton and trion ground- and excited-state populations exist.  The amplitude of the peak at zero mixing energy increases by over an order of magnitude when switching to the HHHH sequence, which we attribute to the allowed exciton quantum-mechanical pathways for this polarization that dominates the nonlinear response \cite{Moody2012}.

\begin{figure}[h]
\centering
\includegraphics[width=0.95\columnwidth]{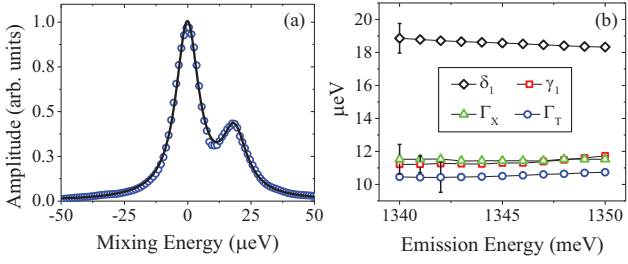}
\caption{(Color online) (a) A slice (points) taken along $\hbar\omega_T$ at $\hbar\omega_t = 1345$ meV and a double (Lorentzian)$^{1/2}$ fit (line) for HVHV polarization. (b) Exciton fine-structure splitting, $\delta_{\textrm{1}}$ (diamonds), $\left|\textrm{H}\right>$--$\left|\textrm{V}\right>$ coherence dephasing rate, $\gamma_{\textrm{1}}$ (squares), and exciton and trion population decay rates, $\Gamma_{\textrm{X}}$ (triangles) and $\Gamma_{\textrm{T}}$ (circles), respectively, are shown as a function of emission energy within the FWHM ground state inhomogeneous distribution.}
\label{fig4}
\end{figure}

To understand how changes in emission energy, and therefore confinement, affect $\delta_{\textrm{1}}$, the dephasing rate of the $\left|\textrm{H}\right>$--$\left|\textrm{V}\right>$ coherence ($\gamma_{\textrm{1}}$) and the population decay rates of the exciton ($\Gamma_{\textrm{X}}$) and trion ($\Gamma_{\textrm{T}}$), we fit slices taken along $\hbar\omega_T$ to either a double (Lorentzian)$^{1/2}$, shown as the solid line in Fig. \ref{fig4}(a) for HVHV at an emission energy of 1345 meV, or a single (Lorentzian)$^{1/2}$ for HHHH.  The fits for VHVH produce similar results for HVHV.  Linewidths and splitting obtained from the fits are shown in Fig. \ref{fig4}(b) for emission energies within the FWHM of the inhomogeneous linewidth.  Estimates for the errors are determined from repeating the experiments.  With increasing (decreasing) emission energy (confinement), $\delta_{\textrm{1}}$ decreases at a rate of 0.1 $\mu$eV$/$meV.  For the ensemble, $\Gamma_{\textrm{X}} \approx 11.6\pm 0.9$ $\mu$eV, $\Gamma_{\textrm{T}} \approx 10.4 \pm 0.8$ $\mu$eV and $\gamma_{\textrm{1}} \approx 11.5 \pm 0.5$ $\mu$eV.  Both $\Gamma_{\textrm{T}}$ and $\gamma_{\textrm{1}}$ increase with increasing emission energy, however the change is within the measured uncertainties.

The population decay rates shown in Fig. \ref{fig4}(b) for the exciton and trion indicate that the radiatively-limited ZPL widths are $\approx 6$ $\mu$eV and $\approx 5$ $\mu$eV, respectively, which are consistent with results obtained from time-integrated FWM studies of samples with similar confinement \cite{Langbein2004}.  The fact that $\gamma_{\textrm{X}} \approx \Gamma_{\textrm{X}}$ and $\gamma_{\textrm{T}} \approx \Gamma_{\textrm{T}}$ indicates that significant pure dephasing exists in these samples even at low temperature.  The linewidths are independent of excitation density, and a temperature dependence shows that the linewidths have reached a low-temperature asymptote at 10 K (data not shown), suggesting that other extrinsic mechanisms are responsible for pure dephasing.  One possible broadening mechanism could be the trapping and escaping of charges in localization sites near the QDs, resulting in a fluctuating quantum-confined Stark shift of the exciton transition energies \cite{Berthelot2006}.  As long as the fluctuations are fast compared to the time between the arrival of the first pulse and the formation of the photon echo, this mechanism could dephase the optical coherences for rephasing one-quantum scans but would have negligible impact on population decay during $T$.

The decrease of $\delta_{\textrm{1}}$ with increasing emission energy is consistent with results in the literature for samples with similar confinement \cite{Langbein2004a,Tartakovskii2004,Ellis2007}, and the results presented here indicate that annealing reduces the rate to 0.1 $\mu$eV$/$meV.  The dephasing rate $\gamma_{\textrm{1}}$ shown in Fig. \ref{fig4}(b) -- equal to $\Gamma_{\textrm{X}}$ within experimental uncertainties -- suggests that not only does a single $\delta_{\textrm{1}}$ characterize the exciton fine-structure splitting at a particular emission energy, but that dephasing of the $\left|\textrm{H}\right>$ and $\left|\textrm{V}\right>$ states are correlated.  To quantify the level of correlation, we can model the nonlinear response using the density matrix formalism for a three-level V-system \cite{Yajima1979,Spivey2008}. \textit{Ignoring population decay}, the dephasing rate of the $\left|\textrm{H}\right>$--$\left|\textrm{V}\right>$ coherence is related to the dephasing rates of the $\left|\textrm{H}\right>$ and $\left|\textrm{V}\right>$ optical coherences by
\begin{equation}
\gamma_{\textrm{1}} = \gamma_{\textrm{H}} + \gamma_{\textrm{V}} - 2 \cdot R \cdot \left(\gamma_{\textrm{H}}\gamma_{\textrm{V}}\right)^{1/2},
\end{equation}
\label{eqn1}
where $\gamma_{\textrm{H}}$ and $\gamma_{\textrm{V}}$ are the dephasing rates of states $\left|\textrm{H}\right>$ and $\left|\textrm{V}\right>$, respectively, and the coefficient $R$ represents the level of correlation between fluctuations in the $\left|\textrm{0}\right>$--$\left|\textrm{H}\right>$ and $\left|\textrm{0}\right>$--$\left|\textrm{V}\right>$ transition energies during scattering events.

\begin{figure}[h]
\centering
\includegraphics[width=0.7\columnwidth]{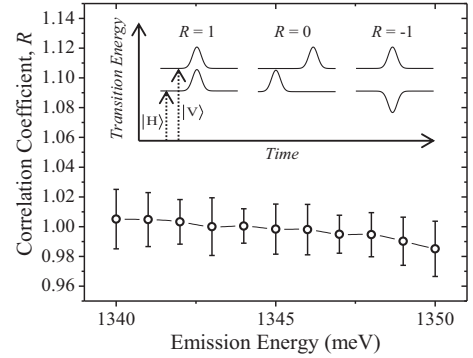}
\caption{Coefficient $R$ representing the level of correlation between fluctuations in the $\left|\textrm{H}\right>$ and $\left|\textrm{V}\right>$ transition energies as a function of emission energy.  Inset: schematic diagram illustrating the effects of correlated ($R = 1$), uncorrelated ($R = 0$) and anti-correlated ($R = -1$) fluctuations as a function of time during a scattering event.}
\label{fig5}
\end{figure}

During a perfectly-correlated scattering event ($R = 1$), the $\left|\textrm{H}\right>$ and $\left|\textrm{V}\right>$ transition energies shift simultaneously and with equal amplitudes, shown in the inset of Fig. \ref{fig5}.  Correlated scattering preserves the $\left|\textrm{H}\right>$--$\left|\textrm{V}\right>$ coherence -- but not the interband coherences of the individual exciton states -- and results in $\gamma_{\textrm{1}} = 0$ for $\gamma_{\textrm{H}} = \gamma_{\textrm{V}}$ and ignoring population relaxation.  For uncorrelated scattering ($R = 0$), the dephasing rate is simply the sum of the individual exciton dephasing rates, while anti-correlated scattering ($R = -1$) -- where the two exciton transitions experience simultaneous equal and opposite transition energy fluctuations -- results in the maximum dephasing rate.  When considering population relaxation, $\gamma_{\textrm{1}} \geq 1/2\cdot\left(\Gamma_{\textrm{X,H}} + \Gamma_{\textrm{X,V}}\right) = \Gamma_{\textrm{X}}$, since for $R = 1$ the $\left|\textrm{H}\right>$--$\left|\textrm{V}\right>$ coherence time is limited by the population lifetimes of the individual exciton states.  Because the experimentally-determined $\gamma_{\textrm{1}}$ is nearly equal to $\Gamma_{\textrm{X}}$, we conclude that the exciton states experience nearly perfectly-correlated scattering and the $\left|\textrm{H}\right>$--$\left|\textrm{V}\right>$ coherence dephasing rate is not further broadened by interband dephasing mechanisms.  $R$ is shown in Fig. \ref{fig5} and is nearly equal to unity for all QDs in the ensemble.  The slight decrease with increasing emission energy originates from the increase in $\gamma_{\textrm{1}}$ since $\gamma_{\textrm{X}}$ and $\Gamma_{\textrm{X}}$ are constant.  This trend likely arises because in QDs with a smaller confinement potential, the $\left|\textrm{H}\right>$ and $\left|\textrm{V}\right>$ wave functions extend into the barrier differently and no longer experience similar scattering events.  The decrease in $R$ could also arise from a distribution of $\delta_{\textrm{1}}$ values existing for higher-energy QDs, which would increase the apparent $\left|\textrm{H}\right>$--$\left|\textrm{V}\right>$ coherence dephasing rate beyond the radiative limit; however this mechanism is unlikely since all QDs are annealed for the same duration and temperature.

\section{Conclusion}
\label{Conclusion}

In conclusion, we measured two-dimensional rephasing zero- and one-quantum spectra of an InAs QD ensemble to study the effects of confinement on the fine-structure splitting and dephasing rate of the non-radiative coherence between the bright exciton states.  We find that the fine-structure splitting decreases with increasing emission energy at a rate of 0.1 $\mu$eV$/$meV.  Fluctuations of the $\left|\textrm{H}\right>$ and $\left|\textrm{V}\right>$ transition energies are nearly perfectly correlated with a minimum correlation coefficient of $R = 0.98$ for the highest-energy QDs, leading to a population lifetime-limited $\left|\textrm{H}\right>$--$\left|\textrm{V}\right>$ coherence dephasing time for essentially all QDs in the ensemble.  Comparison of the population decay rates and ZPL widths reveals that at low temperature and independent of excitation density, pure dephasing processes dephase the optical coherences beyond the radiative limit but do not affect the $\left|\textrm{H}\right>$--$\left|\textrm{V}\right>$ coherence.  One might suspect that with increasing temperature, exciton-phonon interactions could reduce the level of correlation. Investigating the temperature dependence of $R$ could provide a deeper understanding into the mechanisms that govern correlation between of the bright exciton states and is an area of interest for future experiments.

\section*{Acknowledgements}

This work was financially supported by the Chemical Sciences, Geosciences, and Energy Biosciences Division, Office of Basic Energy Science, Office of Science, US Department of Energy, the NSF and the Deutsche Forschungsgemeinschaft.





\bibliographystyle{elsarticle-num}
\bibliography{RefList}







\end{document}